\documentclass[twocolumn,showpacs,preprintnumbers,amsmath,amssymb]{revtex4}
\usepackage{graphicx}
\usepackage{dcolumn}
\usepackage{bm}
\usepackage[dvips]{epsfig}
\usepackage{amsmath}
\usepackage{amssymb}
\def\bea {\begin{eqnarray}}
\def\eea {\end{eqnarray}}

\def\be {\begin{equation}}
\def\ee {\end{equation}}

\begin{document}
\title{Excitation energy and angular momentum dependence of the nuclear level density parameter around A$\approx  $110}
\vspace{.5truecm}
\vspace{.5truecm}

\author{Pratap Roy$^{1,2}$\footnote{Email: roypratap@vecc.gov.in}, S. Mukhopadhyay$^{1,2}$, Mamta Aggarwal$^3$, Deepak Pandit$^{1,2}$, T. K. Rana$^{1,2}$, Samir Kundu$^{1,2}$, T. K. Ghosh$^{1,2}$, K. Banerjee$^{1,2}$, G. Mukherjee$^{1,2}$, S. Manna$^{1,2}$, A. Sen$^{1,2}$, R. Pandey$^{1}$, Debasish Mondal$^1$, S. Pal$^1$, D. Paul$^{1,2}$, K. Atreya$^{1,2}$, and C. Bhattacharya$^{1,2}$ } 
\vspace{.5truecm}
\vspace{.5truecm}
\affiliation{$^1${\em Variable Energy Cyclotron Centre, 1/AF, Bidhan  Nagar, Kolkata~-~700064, India}\\ 
$^2${\em Homi Bhabha National Institute, Training School Complex, Anushaktinagar, Mumbai~-~400094, India}\\
$^3${\em Department of Physics, University of Mumbai, Kalina Campus, Mumbai-400098, India}}
\date{\today}

\begin{abstract}
Neutron kinetic energy spectra in coincidence with low-energy $\gamma $-ray multiplicities have been measured around $A\approx $~110 in the $^{16}$O, $^{20}$Ne~+~$^{93}$Nb reactions in a compound nuclear excitation energy range of $\approx  $~90~-~140 MeV. The excitation energy (temperature) and angular momentum (spin) dependence of the inverse level density parameter $k$ has been investigated by comparing the experimental data with statistical Hauser-Feshbach calculation. In contrast to the available systematic in this mass region, the inverse level density parameter showed an appreciable increase as a function of the excitation energy. The extracted $k$-values at different angular momentum regions, corresponding to different $\gamma $-multiplicities also showed an overall increase with the average nuclear spins. The experimental results have been compared with a microscopic statistical-model calculation and found to be in reasonable agreement with the data. The results provide useful information to understand the variation of nuclear level density at high temperature and spins.         
\end{abstract}


\maketitle
\section{\label{sec1:Intro} Introduction}
Understanding the exact nature of variation of nuclear level density (NLD) as a function of key factors such as excitation energy, angular momentum, iso-spin, shell and collective effects is of particular importance in both nuclear structure and reaction physics. NLD serves as the most critical input in the statistical models used to estimate the reaction cross-sections for various processes ($e.g.$ thermonuclear reactions, fission, evaporation and spallation) in the interdisciplinary areas of nuclear physics, reactor physics and astrophysics. It also acts as a testing ground for different nuclear structure models as well as provides crucial information on the thermodynamic properties of atomic nuclei~\cite{Melby,Giacoppo,Melby2,Agvaan,Toft,Syed,Gutt,PROY,Schil,Kaneko,Chankova,PRiso}. \\
Theoretically, the simplest and most widely used description of level density is given in terms of the non-interacting Fermi gas model ({\scshape FGM})~\cite{Bethe}, 
\begin{equation}
\rho (E)= \frac {1}{12\sqrt{2}\sigma }\frac{\exp{2\sqrt{aE }}}{a^{1/4}E^{5/4}},
\label{eq:eq1}
\end{equation}
where $E$ is the excitation energy and $\sigma $ is the spin cut-off factor. The most important parameter in the {\scshape FGM} description of NLD is the level density (LD) parameter $a$ which is directly related to the density of single-particle states near the Fermi surface. It is well-known that at low energies, the level density is strongly influenced by shell and pairing (odd-even) effects. The pairing effect can be incorporated by shifting the excitation energy on the right-hand side of Eq.~1 by an amount related to the pairing energy ($\Delta $)~\cite{Dlig}. On the other hand, the shell effect can be taken care of through excitation energy-dependent parametrization of the level density parameter as suggested by Ignatyuk {\it et al.}~\cite{Igna},
\begin{equation}
a=\tilde{a}[1-\frac{\Delta S}{U}\{1-\exp(-\gamma U)\}].
\label{eq:eq2}
\end{equation} 
Here $U=E-\Delta $, $\Delta S$ is the ground state shell correction and $\gamma$ is the shell damping parameter. The asymptotic level density parameter $\tilde{a} $, which varies smoothly with the mass number $A$ can be represented as $\tilde{a}= A/k$, where $k$ is called the inverse level density parameter. At high excitation energies or temperatures ($T\gtrsim$~1.5 MeV) the shell effects are fully depleted making $a\approx \tilde{a}$. The angular momentum dependence of NLD described within the Fermi gas picture is given by the Gaussian function exp$(\frac{-(J+1/2)^2}{2\sigma^2})$, where the width of the Gaussian is determined by the temperature ($T$) dependent spin cut-off factor $\sigma $ ($=\frac{IT}{\hbar^2}$ where $I$ is the moment of inertia).\\ 
In the level density formulations based on the simplistic {\scshape FGM}, the level density parameter does not explicitly depend on the excitation energy or angular momentum. However, a number of earlier studies have shown that the parameter $k$ (and thus $a$) depends on excitation energy (temperature)~\cite{Pratap2,Char1,Char2,Nebbia,Hagel,Gonin1,Gonin2,Wada,Fineman,Fabris,Caraley,Chibbi,Nebbia2,Yoshida,Hasse,Shlomo1,Shlomo2,Shlomo3,De}, and angular momentum~\cite{Gohil,Gupta1,Gupta2,Kaushik,Pratap1,Bala2,Mamta1,Mamta2} both, in an intricate manner. Such departures from the {\scshape FGM} may not be surprising as the actual single-particle spectrum of a nucleus is considerably complicated than the simple Fermi gas picture. The experimental data on the spin and excitation-energy dependence of level density thus provide crucial information on the underlying nuclear structure and offer a stringent test for nuclear models. \\
Experimentally, nuclear level density can be computed by different techniques such as the direct counting of nuclear levels~\cite{Kat1,Kat2,Raman,Mishra}, analysis of neutron resonance spacings~\cite{RIPL} and measurement of primary $\gamma $-ray spectra~\cite{Oslo}. However, all these methods are limited to low excitation energies and spins. The major source of knowledge about level densities at higher excitation energies and spins comes from the statistical model analysis of particle-evaporation spectra in heavy-ion fusion reactions~\cite{Char1,Char2,Nebbia,Hagel,Gonin1,Gonin2,Wada,Fineman,Fabris,Caraley,Chibbi,Nebbia2,Yoshida,Hasse}. \\
The analyses of the light-particle evaporation spectra for several medium and heavy nuclei ($A$ $>$150), suggested that the inverse level density parameter should increase systematically with excitation energy, and the energy dependence could be very well represented by a simple linear relationship~\cite{Fineman,Caraley,Char1}
\begin{equation}
k(U)=k_0+\kappa \times (U/A),
\end{equation}
where $k_0$ is the value of $k$ at $U$ or $T$~=~0. The reduction of the level density parameter (an increase of $k$) at higher energies is consistent with the expected fadeout of long-range correlations at higher temperatures. However, the situation for lighter systems has been rather complex since many studies have reported very weak, or no dependence of $k$ on energy~\cite{Chibbi,Nebbia2,Yoshida}. Subsequently, it has been realized that the energy dependence of the LD parameter may depend on the nuclear mass number that can be taken care of by the $A$ dependence of the parameter $\kappa $ (will be called the rate parameter hereafter). Based on the available experimental data a mass number dependent parametrization of $\kappa $ has been suggested by R. J. Charity~\cite{Char1}
\begin{equation}
\kappa =0.00517 \text{exp}(0.0345A).
\label{eq:eq3}
\end{equation} 
However, such a strong mass dependence of the rate parameter is unexpected, and  not supported by theoretical calculations~\cite{Shlomo1,Shlomo2}. Because of the suggested $A$ dependence in Eq.~4, the resultant value of $\kappa $ is minimal for systems with $A<$~150 providing little dependence of $k$ on energy. \\
It should be emphasized here, that experimental data for lighter systems are quite limited, and the uncertainties in the extracted $\kappa $-values are large in this mass region (see $e.g.$ Fig.~9 of Ref.~\cite{Char1}). Moreover, light nuclei have additional complications because of the strong spin dependence of yrast energy ($E_\textrm{yrast}$). This can cause pronounced effects particularly on the predicted spectra of $\alpha $ particles which can remove appreciable amount of angular momentum from the decaying system~\cite{Char1}. Such effects could be much less for neutrons as they tend to remove very little angular momentum and thus are less sensitive to $E_\textrm{yrast}$. However, there is a scarcity of experimental neutron data for lighter nuclei in a wide excitation energy range, and it is demanding to carry out such measurements to understand the excitation energy dependence of the level density parameter in a consistent manner.\\
\indent The level density parameters obtained from the exclusive particle evaporation measurements are average quantities over a range of excitation energies and angular momenta. Angular momentum gated evaporation studies can provide information on the LD parameter at different angular momentum regions. However, the number of such studies are highly limited and, the value of $a$ particularly at high angular momentum is practically unknown for a large number of nuclei. A few attempts have been made in recent years to understand the spin dependence of the level density parameter. Some of these studies showed interesting variations of the LD parameter as a function of angular momentum ($J$). A systematic reduction of $k$ with increasing $<$$J$$>$ in the range of $\approx  $10~-~20 $\hbar $ have been reported in the recent neutron evaporation studies around $A\approx  $~90~-~120~\cite{Kaushik,Pratap1,Bala2}. On the other hand, angular momentum gated $\alpha $-particle evaporation measurement around $A\approx $120, and $J\approx $10~-~20 $\hbar$ showed complex variation of $k$ with angular momentum~\cite{Gupta2}. In contrast to the strong dependence of the LD parameter on the angular momentum reported in the above mentioned works, experimental data for the heavier systems showed less sensitivity of $k$ on $J$~\cite{Gohil,Gupta1}. Some of the experimental data on the angular dependence of $k$ could be successfully explained by the theoretical calculations by considering spin induced deformation and shape phase transitions under the framework of a statistical theory of hot rotating nuclei~\cite{Mamta1,Mamta2}. In view of the observed variation of the level density parameter with $J$ in the earlier works, it will be interesting to extend the study for similar systems especially to higher spin regions.\\    
\indent With the motivations described above we have measured $\gamma $-ray multiplicity gated neutron evaporation spectra from $^{109}$In ($^{16}$O~+~$^{93}$Nb) and $^{113}$Sb ($^{20}$Ne~+~$^{93}$Nb) compound systems. In this paper, we report the average variation of the (inverse) level density parameter as a function of the excitation energy (temperature) and angular momentum for the nuclei around $A\approx  $~110 in the range of $T\approx $~2~-~2.6 MeV and $<$$J$$>\approx $~20~-~36 $\hbar$. The experimental results have been compared with the theoretical calculations performed with a statistical theory of hot rotating nuclei. The present study is expected to provide useful information in understanding the nature of level density at high temperature and spins. \\
\indent The article has been arranged in the following manner. The experimental arrangement has been described in Sec.~\ref{sec2:Expt}. The results have been presented and discussed in Sec.~\ref{sec3:Results}; the excitation energy and temperature dependence of the LD parameter has been presented in Sec.~\ref{sec4:Kex}, and the angular momentum dependence is discussed in Sec.~\ref{sec5:KJ}. Section~\ref{sec6:MSM} gives a brief description of the microscopic MSM calculation. Finally, the present work is summarized in Sec.~\ref{sec7:Sum}.
\begin{figure}
\includegraphics[scale= 0.47]{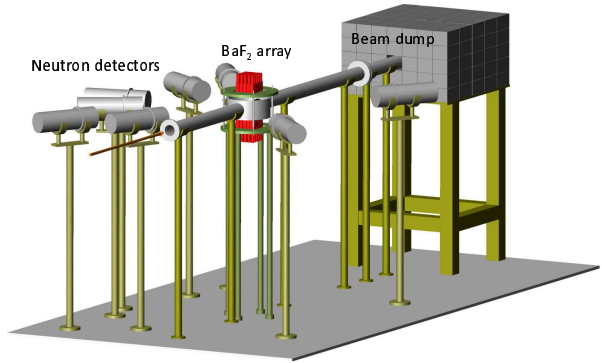}
\caption{(Color online) Schematic representation of the experimental setup.}
\label{fig:fig1}
\end{figure}
\section{\label{sec2:Expt} Experimental Details}
The experiment was performed using $^{16}$O ($E_\textrm{lab}$~=~116, 142 and 160 MeV) and $^{20}$Ne ($E_\textrm{lab}$~=~145 and 180 MeV) ion-beams from the K130 cyclotron at VECC, Kolkata. A self-supporting foil of $^{93}$Nb (thickness $\approx $~2 mg/cm$^2$) was used as the target. The compound nuclei $^{109}$In and $^{113}$Sb were populated in the excitation energy range of $E^\ast_\textrm{CN}\approx  $~90~-~140 MeV. The neutrons emitted during the compound nuclear decay process were detected using eight liquid scintillator based neutron detectors (size: 5-inch $\times $~5-inch) placed at the laboratory angles ($\Theta_\textrm{lab} $) of 45$^\circ $, 60$^\circ $, 75$^\circ $, 90$^\circ $, 105$^\circ $, 120$^\circ $, 135$^\circ $ and 150$^\circ $ at a distance of 2 m from the target. A schematic of the experimental setup used in the present measurement is shown in Fig~1. The neutron kinetic energies were measured using the time-of-flight (TOF) technique. The start trigger for the TOF measurement was generated by detecting the low-energy $\gamma $-rays in a 50-element BaF$_2$ detector array~\cite{Deepak} placed near the target position. In converting the neutron TOF to neutron energy, the prompt $\gamma$ peak in TOF spectrum was used as the time reference. The neutron and $\gamma$ separation was achieved by using both the TOF and pulse shape discrimination methods~\cite{Kaushik2}. The energy-dependent detection efficiency of the neutron detectors were obtained using the Monte Carlo code {\scshape NEFF}~\cite{Klein}. The detector efficiency at low energies (1~-~10 MeV) was also measured experimentally using a $^{252}$Cf source and found to be in good agreement with the {\scshape NEFF} calculation~\cite{Pratap3}.\\ 
The multiplicity of the low-energy $\gamma$ rays was also measured using the BaF$_2$ detector array. The array was split into two blocks of 25 detectors each and was placed on the top and bottom of the thin wall ($\approx $3 mm) reaction chamber (Fig~1). Data from the neutron detectors were recorded in coincidence with $\gamma$ rays of different folds ($F$) which is defined as the number of BaF$_2$ detectors fired simultaneously in an event, and directly related to the populated angular momentum. The angular momentum distributions for different folds were obtained by converting the measured $\gamma$-fold distribution using the Monte Carlo simulation technique based on the {\scshape GEANT3} toolkit, by including real experimental conditions like detector threshold and trigger conditions in the simulation~\cite{Deepak}.\\ 
To keep the background of the neutron detectors at minimum level~\cite{Sujoy}, the beam dump was kept at a distance of $\approx  $~3 m from the target position and was well shielded on all sides with layers of lead and borated paraffin (Fig.~\ref{fig:fig1}). The scattered neutron contribution in the measured neutron spectra was estimated by putting a shadow-bar consisting of 40 cm thick high-density plastic (HDP) and 6 cm Pb blocks in between the target and the detectors. The data were collected in event-by-event mode using a VME based data acquisition system.\\
\section{\label{sec3:Results} Results and Discussions}
\begin{figure}
\centering
\includegraphics[scale= 0.62]{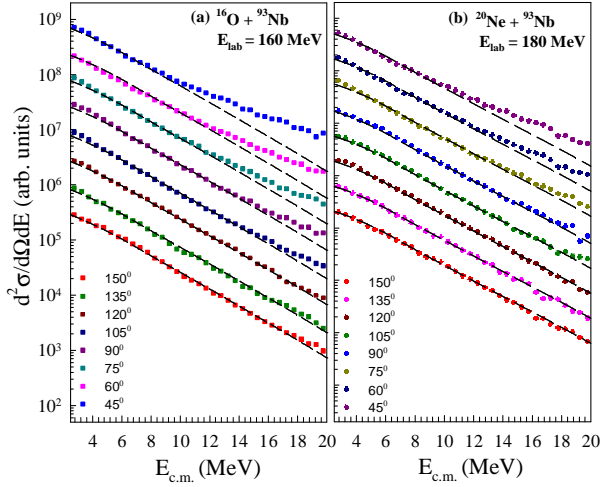}
\caption{(Color online) Experimental inclusive neutron spectra at different angles (symbols) in the c.m. frame (a) for the $^{16}$O~+~$^{93}$Nb reaction at $E_\textrm{lab}$~=~160 MeV and (b) for the $^{20}$Ne~+~$^{93}$Nb at $E_\textrm{lab}$~=~180 MeV. The lines are the prediction of statistical HF calculation. The individual spectrum at different angles has been scaled for better visualization.}
\label{fig:fig2}
\end{figure}

The background-corrected neutron spectra measured at various laboratory angles were transformed into the compound nucleus (CN) center-of-mass (c.m.) frame using the standard Jacobian transformation. The neutron spectra in the c.m. frame measured at different angles at the highest bombarding energies have been shown in Fig.~\ref{fig:fig2}. The experimental spectra have been compared with the statistical Hauser-Feshbach (HF) calculation (shown by the dashed lines in Fig.~\ref{fig:fig2}) performed using the {\scshape CASCADE} computer code~\cite{Puhlhofer}. For the level density, the Reisdorf~\cite{Reis} prescription as presented in Ref.~\cite{Snover} has been used.
\begin{equation}
\rho (E,J) =\frac{2J+1}{12\theta^{3/2}}\sqrt{a}\frac{{\text exp}(2\sqrt{aU})}{U^2}
\label{eq:eq3}
\end{equation}
where
\begin{equation}
U =E-\frac{J(J+1)}{\theta }-\Delta
\label{eq:eq3}
\end{equation}
Here $\Delta $ is the pairing correction which was taken as $\Delta =\delta (12/\sqrt{A})$, where $\delta =$~-1, 0 and 1 for odd-odd, even-odd and even-even nuclei, respectively. Here $\theta =\frac{2I_\textrm{eff}}{\hbar^2 }$; the effective moment of inertia $I_\textrm{eff}$ is related to the rigid body moment of inertia ($I_0$) as $I_\textrm{eff}=I_\textrm{0}(1+\delta_{1}J^2 +\delta_{2}J^4)$~\cite{Cohen}. The quantities $\delta_{1}$ and $\delta_{2}$ known as the deformability coefficients are adjustable input parameters providing a range of choices for the spin dependence of the level density. The default values of $\delta_{1}$ and $\delta_{2}$ are obtained using the rotating liquid drop model~\cite{sierk}. The shell effect in NLD has been incorporated by using Eq.~2; the smoothly varying (asymptotic) level density parameter in Eq.~2 was estimated as $\tilde{a}=A/k $ where $k$ has been treated as a free parameter. 
The transmission coefficients were calculated using the optical model (OM), where the OM parameters were taken from Ref.~\cite{wilmore}. It was observed that the variation in the OM parameters as well as the deformability coefficients had no significant effect on the shape of the calculated neutron evaporation spectra which were mainly decided by the value of the LD parameter. The inverse level density parameter $k$ has been tuned to obtain the best match to the experimental spectra.\\
\indent It can be seen from Fig.~\ref{fig:fig2} that the spectral shapes at the backward angles are almost overlapping, and they agree very well with the statistical model predictions even at the highest incident energies. This indicates that for the present bombarding energies (7~-~10 MeV/$A$) the spectra at back angles are mostly determined by the compound nuclear emission process and any contribution coming from the non-equilibrium processes are small. The existence of the non-equilibrium component is evident in the high-energy tails of the spectra at the forward angles (Fig.~\ref{fig:fig2}). It is interesting to note that qualitatively the non-equilibrium contributions seems to be more for the $^{16}$O~+~$^{93}$Nb reaction than the $^{20}$Ne~+~$^{93}$Nb reaction at similar incident energies.\\
In the present bombarding energy range the spectra at the most backward angle (150$^\circ $) is considered almost free of the non-equilibrium component, and was used for further analysis to understand the excitation energy and spin dependence of the level density parameter.\\ 
\subsection{\label{sec4:Kex} Excitation energy and temperature dependence of $k$}
\begin{figure}
\includegraphics[scale= 0.75]{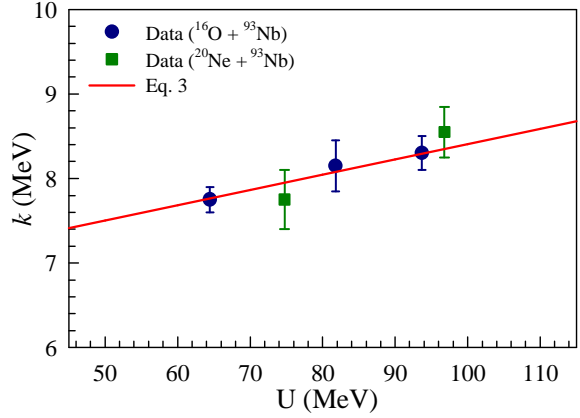}
\caption{(Color online) The excitation energy dependence of the inverse level density parameter $k$.}
\label{fig:fig3}
\end{figure}

\begin{figure}
\includegraphics[scale= 0.72]{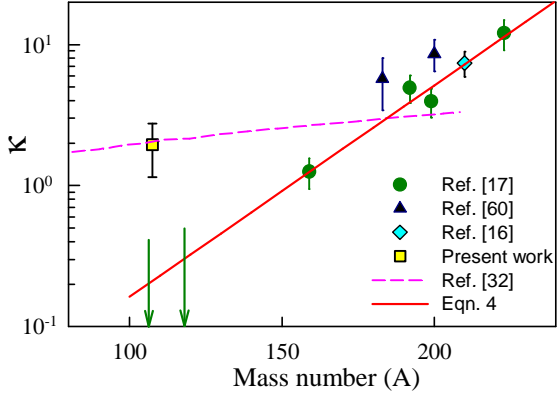}
\caption{(Color online) The values of $\kappa $ as a function of the mass number. The present data is shown by the square. The filled circles and the arrows are regenerated from Ref.~\cite{Char1}. The filled triangles are obtained from the energy dependence of $k$ provided in Ref.~\cite{Banerjee}. The red-continuous line shows the prediction of Eq.~4. The pink-dashed curve shows the $\kappa $ values extracted from the theoretical predictions of Ref.~\cite{Shlomo2}.}
\label{fig:fig4}
\end{figure}

\begin{figure}
\includegraphics[scale= 0.72]{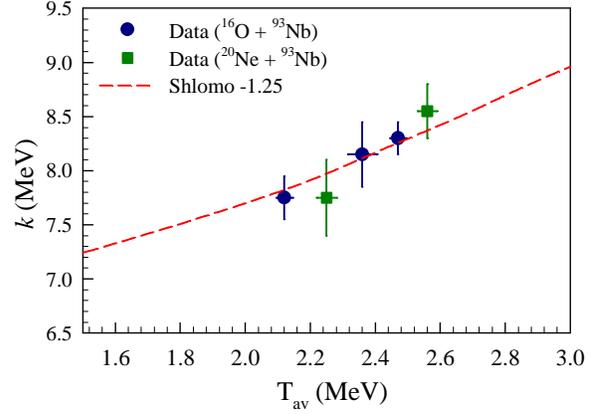}
\caption{(Color online) The temperature dependence of the inverse level density parameter. The experimental data are shown by the symbols. The dashed line shows the prediction of Ref.~\cite{Shlomo2} reduced by a factor of 1.25.}
\label{fig:fig5}
\end{figure}

The back-angle inclusive neutron energy spectra for the $^{16}$O and $^{20}$Ne~+~$^{93}$Nb reactions at different excitation energies were fitted with {\scshape CASCADE} predictions by varying the inverse level density parameter $k$. The optimum values of $k$ corresponding to different excitation energies were extracted by fitting the experimental neutron spectra using the $\chi^2$ minimization technique. The extracted $k$-values as a function of the thermal excitation energy ($U$) has been plotted in Fig.~\ref{fig:fig3}. The thermal excitation energy corresponding to the first stage of the the decay ($i.e.$ after the emission of one neutron) has been estimated from the following relation 
\begin{equation}
U = E^\ast_\textrm{CN}  - <E_{rot}> - S_n - <E_n>	
\label{eq:eq3}
\end{equation}
where $<$E$_{rot}>$ is mean value of the rotational energy for a given $E^\ast_\textrm{CN} $, $S_n$ is the neutron separation energy and $<$$E_n$$>$ is the average kinetic energy of the emitted neutron. It can be observed from Fig.~\ref{fig:fig3} that there is a systematic and appreciable increase of the experimental inverse level density parameter as a function of the thermal excitation energy for the present systems. The experimental data have been fitted with Eq.~3 (shown by the continuous line in Fig.~\ref{fig:fig3}) to obtain the value of the rate parameter ($\kappa $) which comes out to be 1.95~$\pm $~0.75. For the prsent systems the energy dependence of the inverse level desnity parameter is given by $k=6.7+1.95\times (U/A)$.\\
\indent The $\kappa $ value obtained in this work alongwith the previous results~\cite{Pratap2,Char1,Banerjee} have been shown in Fig.~\ref{fig:fig4}. It is interesting to note that the present experimental value of $\kappa $ is significantly higher than the prediction of Eq.~4 (shown by the continuous line in Fig.~\ref{fig:fig4}) for nuclei around $A\approx  $~110. In contrast, the present data agree rather well with the value extracted from the theoretical predictions of Ref.~\cite{Shlomo2} as shown by the dashed line in Fig.~\ref{fig:fig4}. It is evident from the figure that the $\kappa $-values indeed depend on the mass number; however, the dependence may be weaker than the one suggested in Eq.~4. Further experimental data for lighter systems will be useful to get a comprehensive picture of this phenomenon. \\
\indent The dependence of $k$ on nuclear temperature ($T$) has also been investigated and the results are plotted in Fig.~\ref{fig:fig5}. The temperature can be obtained from the thermal energy through the following relation 
\begin{equation}
T= \sqrt{\frac{U}{a}}. 
\label{eq:eq4}
\end{equation}
The temperature obtained in this manner reflects the temperature of the initial stage of the decay cascade. However, the measured neutron spectra contain contributions coming from different stages of the decay. Therefore, it is appropriate to describe the system with an average temperature which is somewhat lower than the one given by Eq.~8. The average or apparent temperatures ($T_\textrm{av} $) corresponding to the different excitation energies were obtained by fitting the experimental spectra with the Maxwellian function $\sqrt{E_{n}}{exp}(-E_{n}/T_\textrm{av})$. The experimental result on the temperature dependence of $k$ (shown by the symbols in Fig.~\ref{fig:fig5}) has been compared with the available theoretical calculation of Shlomo and Natowitz~\cite{Shlomo2} performed under the Thomas-Fermi approach for a nucleus with $A$~=~110.
It should be noted that the calculation of Ref.~\cite{Shlomo2} somewhat over-predicts the absolute values of $k$ obtained in the present work. Therefore, in order to make a meaningful comparison with the data the calculated values of Ref.~\cite{Shlomo2} were scaled down (reduced) by a constant factor of 1.25. It can be observed from Fig.~\ref{fig:fig5} that the observed temperature dependence agrees nicely with the predicted trend of Ref.~\cite{Shlomo2} after the reduction (dashed line in Fig.~\ref{fig:fig5}). It may be mentioned here that the reduction in the value of the level density parameter (the increase of $k$) with temperature can mainly be accounted for, by the temperature dependence of the frequency and momentum dependent effective mass~\cite{Shlomo2}. The frequency dependence of the effective mass, which reflects the effects of correlations, considerably enhances the surface contribution to $a$ at low energies~\cite{Prakash}. However, the effect of correlation dies out with the increase in excitation energy, reducing the value of the level density parameter at higher temperatures.  
\begin{figure*}
\centering
\includegraphics[scale= 0.95]{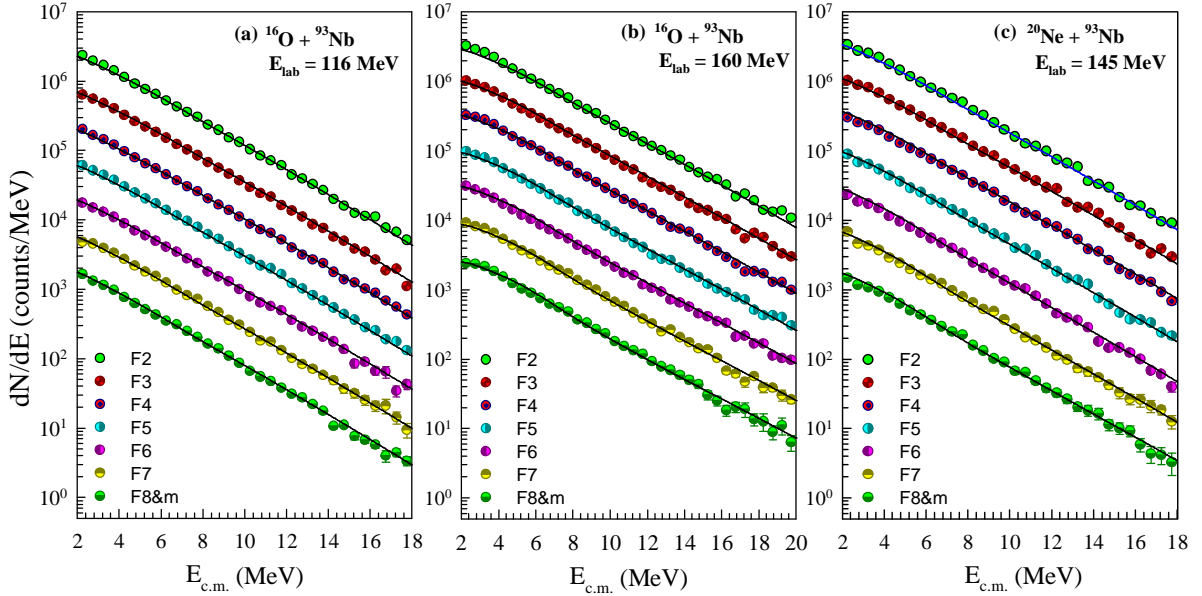}
\caption{(Color online) Experimental neutron spectra (symbols) at different folds for the $^{16}$O~+~$^{93}$Nb reaction at the incident energies of (a) 116 MeV and (b) 160 MeV, and (c) for the $^{20}$Ne~+~$^{93}$Nb at $E_\textrm{lab}$~= 145 MeV. The lines are the prediction of statistical HF calculation. The individual spectrum at at different folds has been scaled for better visualization.}
\label{fig:fig6}
\end{figure*}
\subsection{\label{sec5:KJ} Angular momentum dependence of $k$}
\begin{figure}
\includegraphics[scale= 0.72]{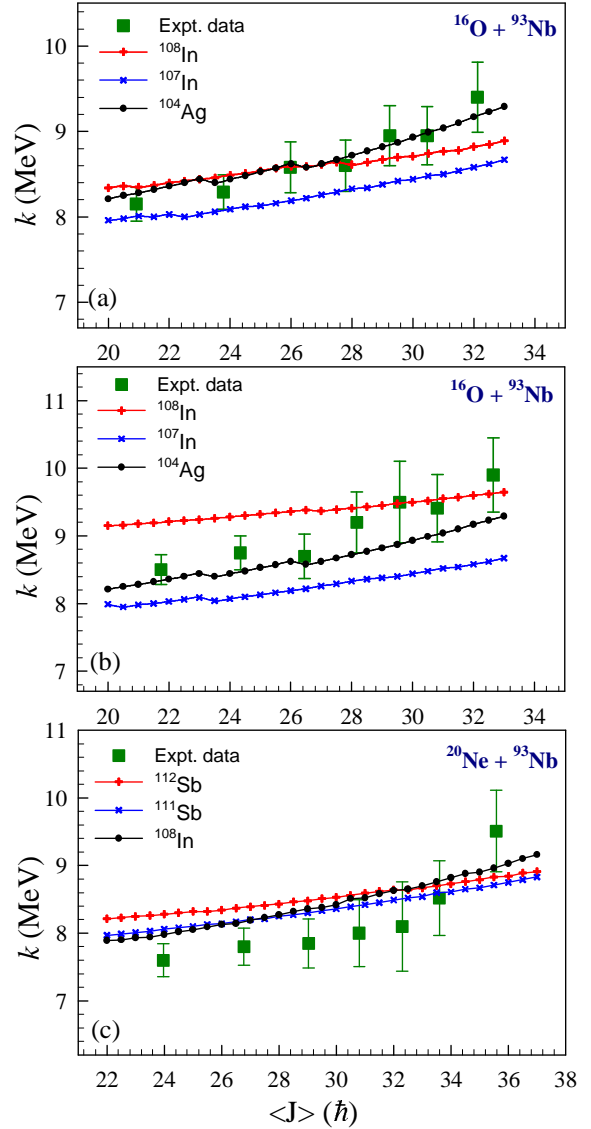}
\caption{(Color online) The experimental $k$-values (symbols) as a function of average angular momenta for the $^{16}$O~+~$^{93}$Nb reaction at the compound nuclear excitation energy ($E^\ast _\textrm{CN}$) of (a) 93.5 and (b) 131 MeV, and (c) for the  $^{20}$Ne~+~$^{93}$Nb reaction at $E^\ast _\textrm{CN}$~=109.5 MeV. The line+symbol plots represent the theoretical predictions for the three most significant daughter nuclei obtained within the statistical model of hot-rotating nuclei (see the text).}
\label{fig:fig7}
\end{figure}

The experimental neutron energy spectra for different $\gamma $-folds ($F $), corresponding to different angular momentum regions were extracted and compared with the theoretical {\scshape CASCADE} calculations as shown in Fig.~\ref{fig:fig6}. As mentioned, the angular momentum distributions corresponding to different $\gamma $-folds have been extracted by using the {\scshape GEANT3} based simulation technique described in Ref.~\cite{Deepak} in detail. The theoretical neutron energy spectra were calculated using {\scshape CASCADE}, with the extracted angular momentum distributions for different folds as inputs. The best-fit values of the inverse level density parameter as obtained from the theoretical fits to the neutron spectra, for different folds, are given in Table I.\\ 
The $k$-values as a function of the mean angular momentum in the daughter nuclei have been plotted in Fig.~\ref{fig:fig7}. It is observed that the experimental $k$-values increases as a function of the mean angular momentum for both the $^{16}$O~+~$^{93}$Nb and $^{20}$Ne~+~$^{93}$Nb reactions in the measured angular momentum range of $<$$J$$>\approx $~20~-~36 $\hbar$. The experimental results have been compared with the microscopic calculations performed using the statistical model of hot rotating nuclei~\cite{Mamta1,Mamta2,Mamta3}, described briefly in the following section (Sec.~\ref{sec6:MSM}). It should be mentioned that the experimental neutron spectra contain contributions coming from the different stages of the decay leading to different residual nuclei. Therefore, the extracted level density parameters do not strictly correspond to any specific daughter nucleus, rather they represent the average value for nuclei in the given mass region. In order to compare with the experimental data the theoretical calculations were performed for the three most significant daughter nuclei corresponding to the 1$n$, 2$n$ and $\alpha n$ decay channels ($i.e.$ $^{108}$In, $^{107}$In and $^{104}$Ag for the $^{16}$O~+~$^{93}$Nb reaction and $^{112}$Sb, $^{111}$Sb and $^{108}$In for the $^{20}$Ne~+~$^{93}$Nb reaction). The outcome of this calculation has been shown by the line plus symbol plots in Fig.~\ref{fig:fig7}. The experimental data on the average agree reasonably well with the predicted values. 
\begin{center}
\begin{table}[ht]
\caption{\label{tab:table1} Average angular momenta in the residual nuclei and the extracted inverse level density parameters for different $\gamma$-folds.}
{\small
\hfill{}
\begin{ruledtabular}
\begin{tabular}{c c c c c}

System & $E_\textrm{lab}$   & Fold& $< J >$  & ($k$ ) \\
 (CN)&  ($E^\ast _\textrm{CN}$) & &  ($\hbar$) &  (MeV) \\
  &  (MeV) & &   &   \\
 \hline
					&  		& 2	& 20.9	& 8.1$\pm$0.2 \\
					& 	 	& 3 & 23.8	& 8.3$\pm$0.2 \\
$^{16}$O+$^{93}$Nb	& 116	& 4	& 26.0	& 8.6$\pm$0.3  \\
					& 		& 5	& 27.8	& 8.6$\pm$0.3 \\
($^{109}$In)		&(93.5)	& 6 & 29.2	& 8.9$\pm$0.3 \\
					&  		& 7	& 30.5	& 9.0$\pm$0.4  \\
					&  		&$\geq$8	&32.1	& 9.4$\pm$0.4  \\
 \hline
 					&  		& 2	& 21.7	& 8.5$\pm$0.2 \\
					& 	 	& 3 & 24.3	& 8.7$\pm$0.2 \\
$^{16}$O+$^{93}$Nb	& 160	& 4	& 26.4	& 8.7$\pm$0.3  \\
					& 		& 5	& 28.2	& 9.2$\pm$0.4 \\
($^{109}$In)		&(131.0)	& 6 & 29.6	& 9.5$\pm$0.6 \\
					&  		& 7	& 30.8	& 9.4$\pm$0.5  \\
					&  		&$\geq$8	&32.6	& 9.9$\pm$0.5  \\
\hline
 					&  		& 2	& 24.0	& 7.6$\pm$0.2 \\
					& 	 	& 3 & 26.8	& 7.8$\pm$0.3 \\
$^{20}$Ne+$^{93}$Nb	& 145	& 4	& 29.0	& 7.8$\pm$0.4  \\
					& 		& 5	& 30.8	& 8.0$\pm$0.5 \\
($^{113}$Sb)		&(109.5)	& 6 & 32.3	& 8.1$\pm$0.7 \\
					&  		& 7	& 33.6	& 8.5$\pm$0.5  \\
					&  		&$\geq$8	&35.6	& 9.5$\pm$0.6  \\
 
\end{tabular}
\end{ruledtabular}}
\hfill{}
\end{table}
\end{center}
\subsection{\label{sec6:MSM} Microscopic calculation and comparison with the data}
To investigate the observed energy and angular momentum dependence of the level density parameter theoretically, a microscopic calculation has been performed within the theoretical framework that involves the statistical theory~\cite{Mamta1,Mamta2,Mamta2,MRSPRC}, and the Macroscopic-Microscopic approach using the triaxially deformed Nilsson-Strutinsky model~\cite{MASHA,MAPL}. In this model, the excited compound nuclei are described as the thermodynamical system of fermions incorporating their collective and non-collective rotational degrees of freedom. The basic ingredient of the theory is a suitable shell-model level scheme for various deformations, which is generated by assuming that the nucleons move in a deformed oscillator potential of the Nilsson Hamiltonian, diagonalized with cylindrical basis states~\cite{Shan,Eisen} with the Hill-Wheeler~\cite{Hill} deformation parameters. The levels up to N = 11 shells of the Nilsson model with Seeger parameters~\cite{Seeger} are used. \\
The equilibrium deformation ($\beta$) and shape ($\gamma$) of the nucleus have been determined by minimizing the appropriate free energy $F=E-TS$~\cite{GOODF}. The minima of $F$ have been traced with respect to the intrinsic shape parameters $\beta$ and $\gamma$ which also describe the orientation of the nucleus with respect to its rotation axis. The total energy ($E$) and entropy ($S$) which are functions of particle number, deformation and shape alongwith the orientation with respect to rotation axis are computed within this microscopic statistical model (MSM).\\
Having the single-particle ($s.p.$) level scheme, the occupation probabilities of these $s.p.$ levels are calculated at different temperatures ($T$) by following the Fermi distribution function. The corresponding excitation energies ($U$) are then extracted by adding the single-particle energies of the occupied levels and subtracting the ground state energy from it~\cite{Mamta2,Mamta3}. Subsequently, the level density parameter is obtained by using the Fermi gas formula $a=U/T^2$, and the inverse level density parameter $k$ is evaluated as $k=A/a$. The total level density ($\rho $) has been calculated using Eq.~5 with the MSM computed level density parameters. The microscopic model used in the present work have been adequately described in Refs.~\cite{Mamta2,Mamta3,MAPL}, and the reader may refer to these articles for the details of the formalism. \\
\begin{figure}
\centering
\includegraphics[scale= 0.70]{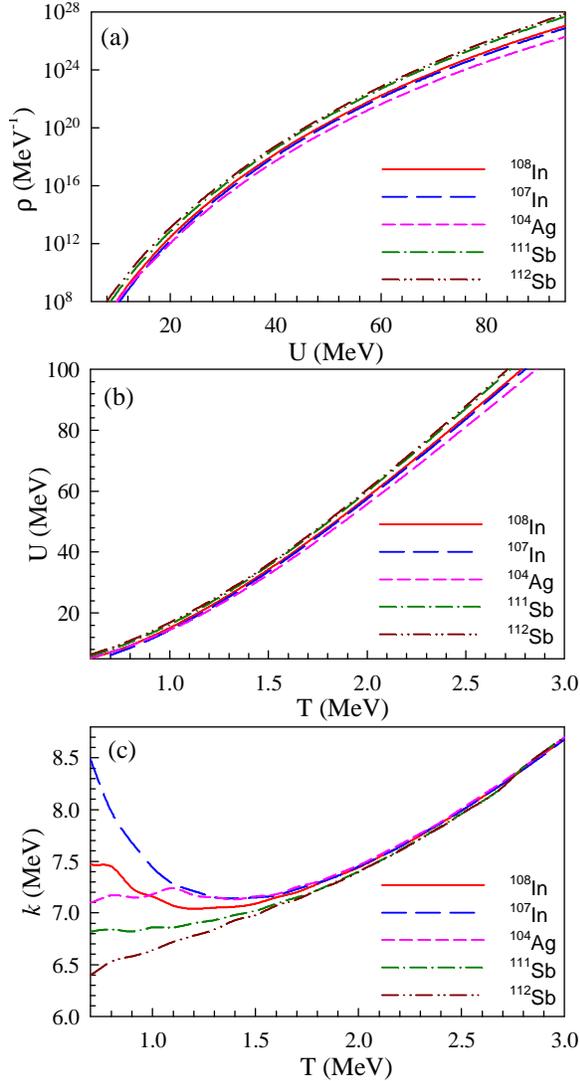}
\caption{(Color online) The theoretical results for different nuclei obtained within the microscopic statistical model (see the text). (a) The energy dependence of the total level density. The temperature dependence of the (b) thermal energy and (c) the inverse level density parameter. }
\label{fig:fig8}
\end{figure}
The nuclear level density, the temperature dependence of excitation energy and the inverse LD parameter obtained from the MSM calculation for the residual nuclei $^{108}$In, $^{107}$In, $^{104}$Ag, $^{112}$Sb and $^{111}$Sb in the excitation energy range covered experimentally in the present work are shown in Fig.~\ref{fig:fig8}. The excitation energy increases with temperature approximately as the Fermi gas relation $U = aT^2$ (Fig.~\ref{fig:fig8}(b)). The shell effects are prominent at low $T$, which is visible in the fluctuations in the $k$ values, as shown in Fig.~\ref{fig:fig8}(c). After the shell effects get quenched ( for $T\gtrsim$~1.5 MeV), the inverse level density parameter $k$ increases smoothly with temperature for all the systems (Fig.~\ref{fig:fig8}(c)); the trend agrees reasonably with the average variation of $k$ observed in the present experimental data as shown in Fig.~\ref{fig:fig5}. The predicted results also match with the trend shown in Refs.~\cite{Shlomo2,BK,Prajapati}. The nuclear level density $\rho(U)$ as shown in Fig.~\ref{fig:fig8}(a) increases rapidly with $U$ (and $T$) and eventually slows down at higher energies where $k$ increases with temperature.\\  
The angular momentum dependence of $k$ has also been investigated within the MSM. The $k$-values at a given initial temperature corresponding to the experimental excitation energies for the different systems have been evaluated at different angular momenta in the range of $J=$ 20~-~40 $\hbar $. The results are shown in Fig.~\ref{fig:fig7} for the different daughter nuclei. The predicted angular momentum dependence for the nuclei under consideration matches with the average variation of $k$ with $<$$J$$>$ observed experimentally in the present work. For a given energy of the system, even though the available thermal energy is reduced by the increasing the rotational energy at higher spins the $k$-values are observed to increase with $J$ indicating a relative reduction of the level density at higher angular momentum.   \\
\section{\label{sec7:Sum} Summary and Conclusions}
The neutron kinetic energy spectra emitted from the excited $^{109}$In and $^{113}$Sb compound systems have been measured in the excitation energy range of $E^\ast_\textrm{CN}\approx $~90~-~140 MeV. The multiplicity of low-energy $\gamma $ rays was also measured to estimate the populated angular momentum. The theoretical analysis of the neutron spectra was performed within the statistical Hauser-Feshbach formalism to investigate the excitation energy (temperature) and angular momentum dependence of the level density parameter for the nuclei around $A=$~110. The experimental data clearly show that the level density parameter $a$ decreases with the increase in excitation energy (temperature) as well as angular momentum. The energy dependence of the level density parameter for the present systems could be expressed as $\tilde{a}(U)=\frac {A}{6.7+1.95(U/A)}$. The observed variation of the LD parameter is in contrast to the available systematic in this mass region; however, the trend matches nicely with the theoretical calculations. It would be interesting to carry out further experimental investigations in this mass region to get a more comprehensive picture of the observed phenomenon.\\
The angular momentum dependence of $a$ has been investigated from the analysis of $\gamma $-ray multiplicity gated neutron spectra, and it is observed that for a given initial temperature, the $k$-values increase with $<$$J$$>$. The results indicate that for a given initial energy, there is a relative reduction of the level density at higher angular momentum. The experimental results were compared with the microscopic statistical-model calculation and found to be in good agreement.\\ 
\section{Acknowledgments}
The authors would like to acknowledge the VECC Cyclotron operators for the smooth running of the accelerator during the experiment. We are also thankful to J. K. Meena, A. K. Saha, J. K. Sahoo and R. M. Saha for their help during the experimental setup.\\

\end{document}